\documentclass{ws-ijmpe}

\begin{document}

\markboth{Ga\"el Renault \it{et al.}}{The Laser of the ALICE Time Projection Chamber}

\catchline{}{}{}{}{}
%%%%%%%%%%%%%%%%%%%%%%%%%%%%%%%%%%%%%%%%%%%%%%%%%%%%%%%%%%%%%%%%%%%%

\title{THE LASER OF THE ALICE TIME PROJECTION CHAMBER}

\author{G. RENAULT, B.S. NIELSEN, J. WESTERGAARD AND J.J. GAARDH\O JE \\
FOR THE ALICE COLLABORATION} 

\address{Niels Bohr Institute, Blegdamsvej 17, \\
2100 Copenhagen, Denmark \\
renault@nbi.dk}

\maketitle

\begin{history}
\received{(received date)}
\revised{(revised date)}
%\accepted{(Day Month Year)}
%\comby{(xxxxxxxxxx)}
\end{history}

\begin{abstract}
The large TPC ($95~\mathrm{m}^3$) of the ALICE detector at the CERN LHC
was commissioned in summer 2006. The first tracks were observed both
from the cosmic ray muons and from the laser rays injected into the TPC.
In this article the basic principles of operating the $266~\mathrm{nm}$
lasers are presented, showing the installation
and adjustment of the optical system and describing the control system.
To generate the laser tracks, a wide laser beam is split into several hundred
narrow beams by fixed micro-mirrors at stable and known positions throughout
the TPC. In the drift volume, these narrow beams generate straight
tracks
%by two-photon ionization
at many angles. 
%Other electrons are
%emitted by the photo-electric effect when the laser beam hits metallic
%surfaces such as the central electrode, the aluminized mylar strips
%of the electric field cage, wires and pads in the readout chambers.
Here we describe the generation of the first tracks
and compare them with simulations.
\end{abstract}

\section{Introduction}

The ALICE experiment will study heavy ion collisions at LHC.
The main tracking detector of the ALICE experiment is the TPC\cite{TDR}.
The aim of the laser system is to generate straight tracks,
similar to ionizing particle tracks,
at known positions in the drift volume of the TPC. 
These tracks are generated by the two-photon
ionization of the drift gas
by a pulsed UV laser beam with a wavelength of $266~\mathrm{nm}$.
Other electrons are emitted
by photoelectric effect when the laser beam hits metallic surfaces
such as the central electrode, the aluminized mylar
strips of the electric field cage, or wires and pads of the readout system.
After readout and track reconstruction using the TPC detector,
distorsions related to ExB effects and mechanical
misalignment will be measured and corrected using these tracks.
The spatial and temporal variations of
the drift velocity due to the drift field will be measured within
a relative error of $10^{-4}$ and used as calibration data
in the physics analysis.

\section{The TPC laser system}

\begin{figure}[hbt!]
\centerline{\psfig{file=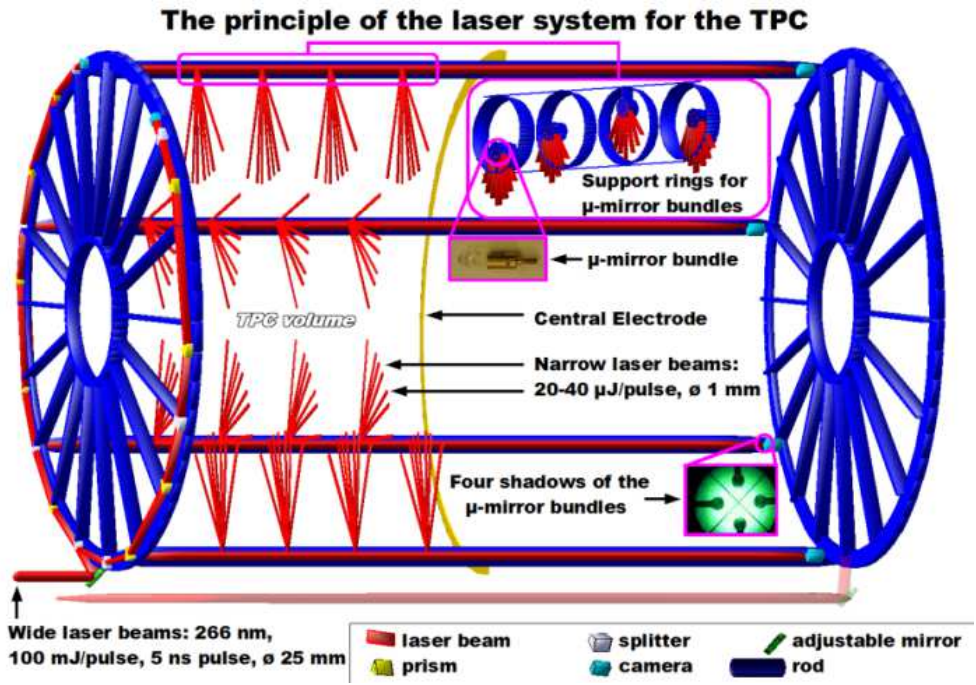,width=13cm}}% or 10cm
\vspace*{8pt}
\caption{The principle of the laser system for the TPC.
A 25 mm wide and 5 ns long pulse hits an adjustable mirror
and is guided inside the TPC. In the laser rode, the beam is split
into one-millimeter diameter rays to simulate ionizing particle tracks
in the TPC drift volume.}
\label{principle3D}
\end{figure}

\begin{figure}[hbt!]
\centerline{\psfig{file=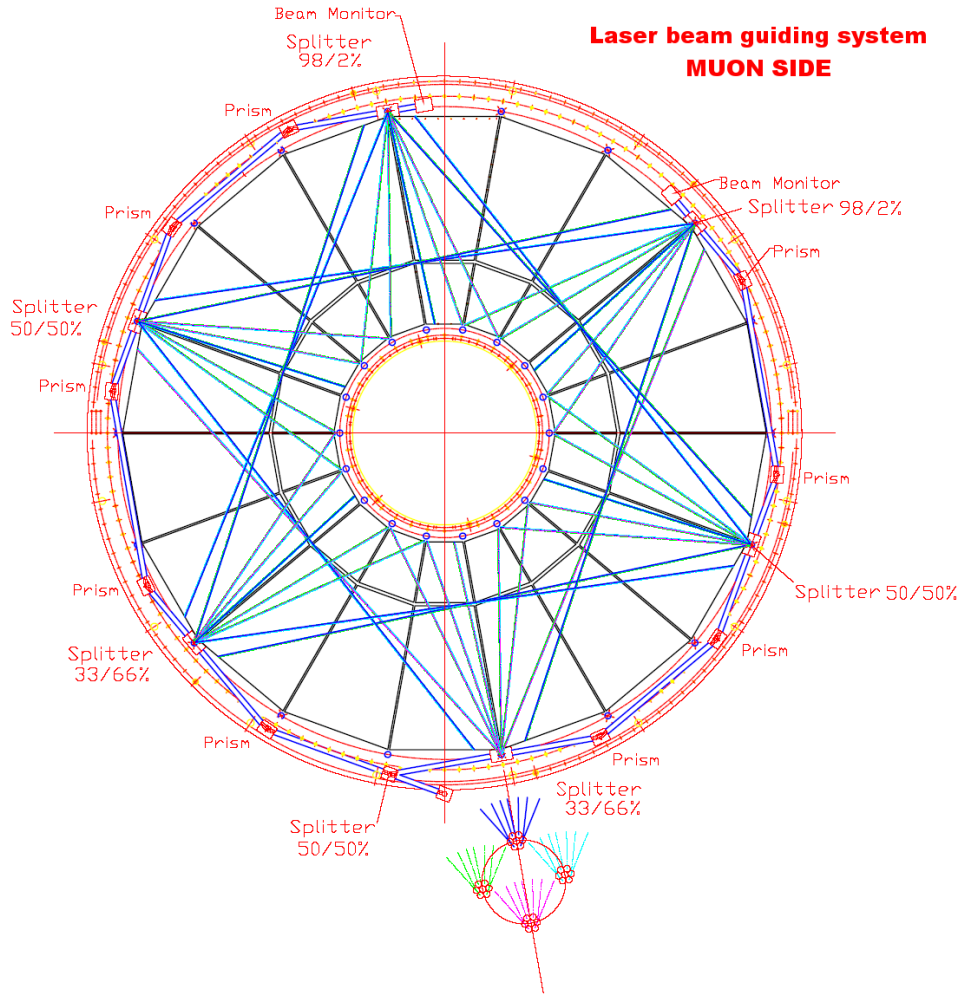,width=13cm}}% or 10cm
\vspace*{8pt}
\caption{The design of the laser beam guiding system. In each of the $6$ rods, $4$ micro-mirror bundles reflect a part of the laser beam
into $7$ one-millimeter diameter beams that enter the TPC volume.}
\label{laser-tracks-designed}
\end{figure}

%\subsection{The principle}
The calibration system (Fig. \ref{principle3D})
is composed of a static optical system
with a few adjustable parts. The static optics is composed of beam
splitters, mirrors and bending prisms guiding the laser beam
outside the TPC field cage before it enters the TPC volume.
The guiding system ends with cameras that take pictures
of the laser beam in order to monitor the position and the beam intensity.  
The adjustable part is mainly composed of remotely adjustable mirrors
that will guide the beam into the static optics system.
In order to generate multiple tracks in the TPC, the laser beam
goes through several steps. 
First, a $25 \mathrm{mm}$ diameter laser beam is divided,
at one end, into $6$ by beam
splitters before it enters the TPC. Each beam enters
the TPC in a hollow rod, goes through the central electrode and, for
monitoring purposes is detected by a camera located at the far end
of the TPC. In each rod, $4$ micro-mirror bundles, placed at fixed positions
along the length of the rod, reflect a part of the laser beam
into $7$ one-millimeter diameter beams (20-40 $\mu$J/pulse)
that enter the TPC volume. A second laser generates similar
rays in the second half of the TPC.
Thus, on the whole $336$ laser beams (Fig. \ref{laser-tracks-designed})
are created inside the TPC volume\cite{LASERCALIB2002}.

%\subsection{The two lasers}

A laser from Spectron Laser Systems Ltd
(model SL805-UPG) is set up in the laser hut. 
A second laser from Ekspla company\cite{EKSPLA} (model NL313)
will be set up in the experimental area.
These lasers deliver $266~\mathrm{nm}$ wavelength pulses of $5~\mathrm{ns}$ 
duration at a repetition rate of $10~\mathrm{Hz}$ and the energy
of a pulse is $100~\mathrm{mJ}$
at the entrance of the end cap.
Each laser device has two RS232 connections for control
and monitoring purposes.
They are converted into a TCP/IP connection
with an interface from Digi International\cite{DIGI} company in order
to be able to control the lasers over long distances.    
Each laser is controled by C++ serial port
drivers included in two separate DIM\cite{DIM} servers running on Windows XP.
The servers communicate with DIM clients included in the user interfaces.
The two lasers can be used
in parallel, one for each end cap, or one for both
end caps, the other one being kept as a backup \cite{LASERCALIB2002,LASERDCS}.

\begin{figure}[hbt!]
\centerline{\psfig{file=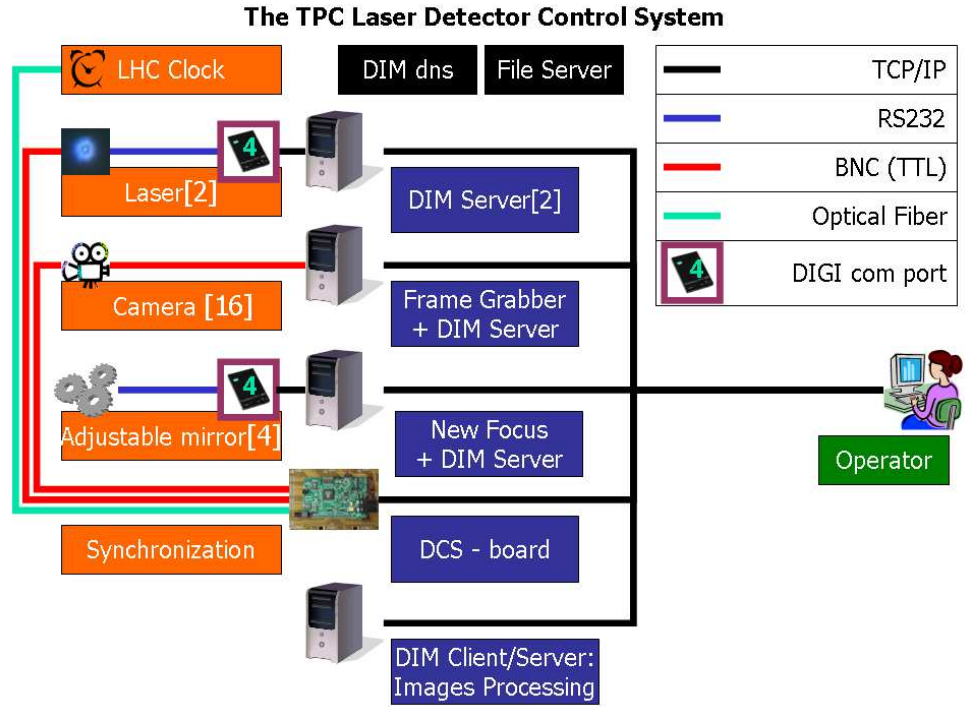,width=13cm}}% or 10cm
\vspace*{8pt}
\caption{The overview of the TPC Laser Detector Control System.}
\label{dcs-schema}
\end{figure}

%\subsection{Adjustable mirrors}
Two adjustable mirrors are used to align the laser
beam in the guiding system on the TPC endplates.
They are controlled by New Focus\cite{NEWFOCUS} pico-motors.
All pico-motors are controlled by C++ serial port drivers included
in a DIM server.
The RS232 communication protocol is converted into TCP/IP,
with an interface from Digi International\cite{DIGI} company,
to be able to control the pico-motors over long distances.

%\subsection{Cameras}      % \subsection*{}
Cameras (up to 16) will monitor the laser
beams. At each end cap one camera is placed behind
the movable mirror. Two others can be placed at
the end of the two $180^{\circ}$ bends.
Moreover, 12 extra cameras are at the end of each rod.
All these cameras will take pictures of the laser beam
in order to monitor its intensity and its position.
Images are acquired with a frame grabber card (and an extension card)
from Imaging Development Systems\cite{IDS} (IDS), model FALCON-quattro. 

%\subsection{The synchronization board}
An electronic board
will synchronize laser pulses and cameras with the LHC clock.
This board was developed for the Detector Control System
of the TPC and TRD detector in ALICE\cite{TDR,TRD}.

\section{The control system}

The TPC laser Detector Control System (Fig. \ref{dcs-schema}) is
used to control and monitor the laser beams from
the control room while sending pulses through the
ALICE TPC. Images of the laser spots recorded by
cameras will be processed in order to move adjustable
mirrors to align laser beams inside the rods.
The DCS board will be configured to synchronize the
laser pulse emission and the image capture (by cameras
and frame grabber cards) with the LHC clock.
All sub-systems are controlled by a common Supervisory
Control And Data Acquisition (PVSS, Process
Visualization and Steering System\cite{PVSS} from ETM Professional
Control company) inside the Joint Controls
Project framework\cite{framework}, giving a stable user interface.

\begin{figure}[hbt!]
\centerline{\psfig{file=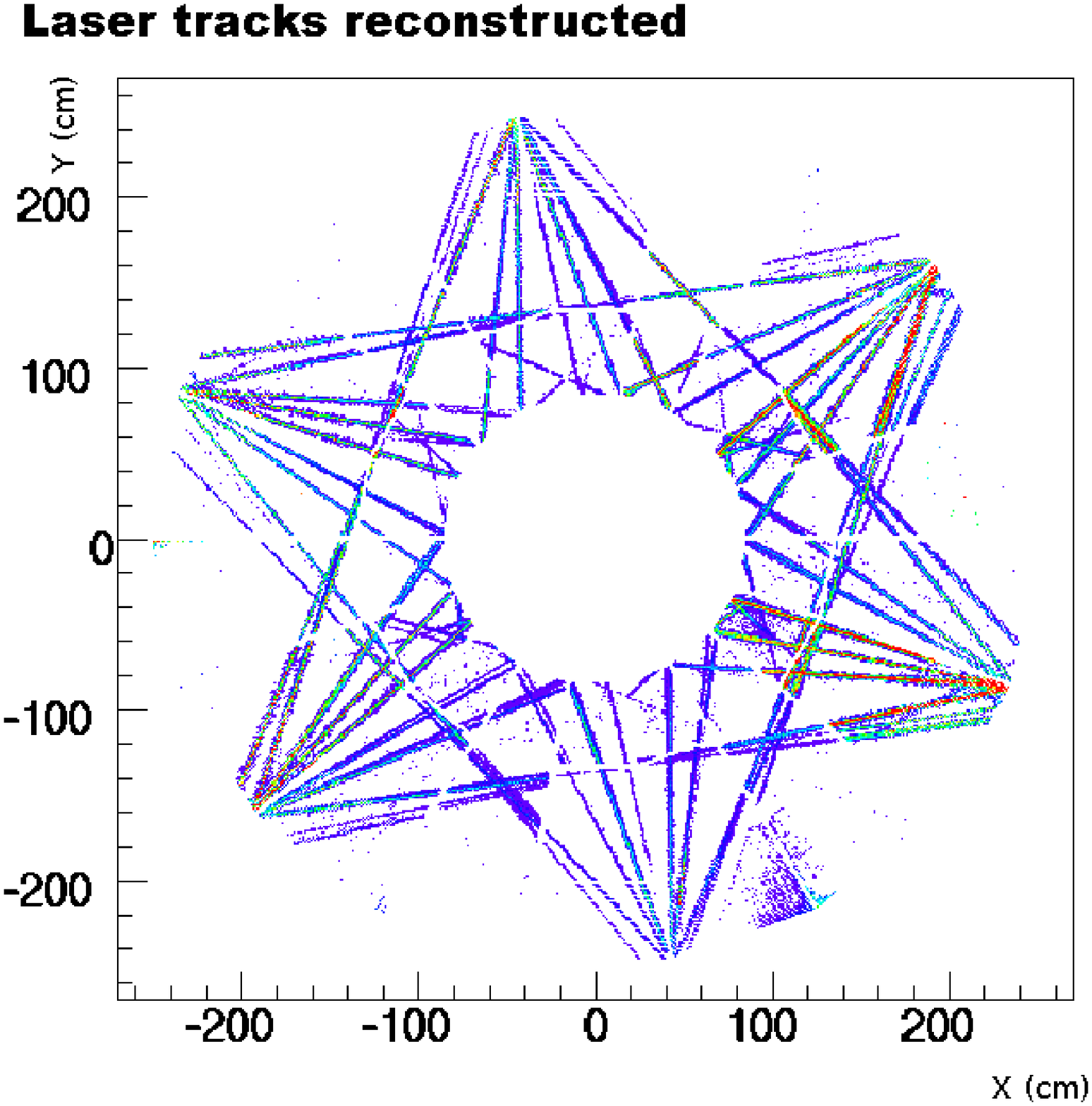,width=13cm}}% or 10cm
\vspace*{8pt}
\caption{The laser tracks reconstructed in the side A of the TPC.}
\label{reco-laser}
\end{figure}

\section{The data analysis}

The TPC reconstruction software was tested and optimized on simulated
laser events, where the straight
laser tracks in the TPC detector were approximated
by high transverse momentum muons emitted from the micro-mirrors,
using the AliRoot\cite{aliroot} framework. The reconstruction algorithms of
the TPC tracks were used to reconstruct the simulated muons.
Last summer, data was recorded in two opposite sectors at the same time during
the TPC commissioning. While the data analysis is still going on, preliminary
results on the reconstructed laser tracks (Fig. \ref{reco-laser})
are in agreement with the simulated tracks.

\section{Conclusion}

The ALICE TPC has been commissioned with
both cosmic muon tracks and straight track
events generated by the laser system. Further
data analysis will make it possible to study and
correct for chamber alignment and drift distortion
effects before the installation of the TPC in
the ALICE experiment. A second laser and the
full Detector Control System will be installed in
the next months and will be used to monitor
and correct distortions for the duration of the
ALICE experiment.

%%%%%%%%%%%%%%%%%%%%%%%%%%%%%%%%%%%%%%%%%%%%%%%%%%%%%%%%%%%%%%%%%%%
%%% 			BIBLIOGRAPHY				%%%
%%%%%%%%%%%%%%%%%%%%%%%%%%%%%%%%%%%%%%%%%%%%%%%%%%%%%%%%%%%%%%%%%%%

\end{document}